\documentclass[eqsecnum,showpacs,preprintnumbers,amsmath,amssymb]{revtex4}
\usepackage{graphicx}
\begin{document}
\preprint{IMAFF-RCA-04-04}
\title{Phantom Thermodynamics}
\author{Pedro F. Gonz\'{a}lez-D\'{\i}az$^{1}$ and Carmen L. Sig\"{u}enza}
\affiliation{$^1$Colina de los Chopos, Centro de F\'{\i}sica ``Miguel
A. Catal\'{a}n'', Instituto de Matem\'{a}ticas y F\'{\i}sica Fundamental,\\
Consejo Superior de Investigaciones Cient\'{\i}ficas, Serrano 121,
28006 Madrid (SPAIN)}
\date{\today}
\begin{abstract}
This paper deals with the thermodynamic properties of a phantom
field in a flat Friedmann-Robertson-Walker universe. General
expressions for the temperature and entropy of a general
dark-energy field with equation of state $p=\omega\rho$ are
derived from which we have deduced that, whereas the temperature
of a cosmic phantom fluid ($\omega<-1$) is definite negative, its
entropy is always positive. We interpret that result in terms of
the intrinsic quantum nature of the phantom field and apply it to
(i) attain a consistent explanation for some recent results
concerning the evolution of black holes which,induced by accreting
phantom energy, gradually loss their mass to finally vanish
exactly at the big rip, and (ii) introduce the concept of
cosmological information and its relation with life and the
anthropic principle. Some quantum statistical-thermodynamic
properties of the quantum quantum field are also considered that
include a generalized Wien law and the prediction of some novel
phenomena such as the stimulated absorption of phantom energy and
the anti-laser effect.
\end{abstract}

\pacs{98.80.-k, 98.80.Hw}

\maketitle

\section{Introduction}

Phantom energy remains for the moment a theoretical possibility
[1] with potential application to describe super-accelerated
evolution in both the primordial and late universe. But it is not
just that. Concerning late evolution of the universe, the
possibility that the equation of state governing current cosmic
evolution would correspond to phantom energy is still not at all
excluded [2]. Actually, if the accelerating expansion of the
universe turned out to be not due to the existence of a positive
cosmological constant, then presently supplied cosmic data appear
to favor phantom energy over quintessence models for positive
internal energy fields [3]. From a theoretical point of view, the
study of phantom energy is by itself a very interesting subject.
It actually bears some resemblance with the study of black holes
in the early seventies by its intrinsic interest and in what then
black holes were considered to be on the borderline of being or
not actual objects. It is therefore not surprising that phantom
energy had recently received a great deal of attention [4].

Phantom energy can be defined in all present models for dark
energy by simply introducing in such models a spontaneous
violation of the dominant energy condition; that is, if e.g. we
consider an equation of state $p=\omega\rho$, then the phantom
appears when we enforce $p+\rho <0$, which in turn implies that
the parameter $\omega<-1$. Thus, in order to check the occurrence
of phantom energy in the universe, it suffices to determine what
the equation of state of the universe is. Actually cosmic phantom
energy keeps a fairly large set of rather weird properties,
including [4] the above-mentioned violation of the dominant energy
condition, $p+\rho<0$, naive superluminal sound speed, increasing
energy density with cosmic time, and ultimately the emergence in a
finite time in the future of what has been dubbed big rip [5],
which is a true curvature singularity where all existing particles
are coherently ripped apart. The latter prediction actually was
already advanced in 1986 by Barrow, Galloway and Tipler [5]. In
this paper we shall add some extra weirdness concerning the
thermodynamic properties of phantom energy; more precisely, we
show that phantom energy is characterized by a negative
temperature and that this implies the phantom to be quantized,
even though it is the largest possible system. This overall
quantum nature of the cosmic phantom fluid comes about as a
consequence from preserving the weak energy condition, and can be
shown by either taking the phantom stuff to be made up of
super-light axions [7], or defining the phantom spacetime to be
Euclidean.

Thus, the aim of the present paper is at studying the
thermodynamic properties of cosmic phantom energy. The influence
that the presence of black holes may have on the phantom field is
also explored. The main results are: (i) we confirm and interpret
the result obtained by Babichev, Dokuchaev and Eroshenko [8] that
all black holes in the universe loss their mass to vanish at the
big rip; (ii) since the internal phantom energy is negative,
whereas the phantom temperature is definite negative, and hence
hotter than any other sources in the universe [9], its entropy is
always positive, even though holding of the second law is not
guaranteed by quantum-mechanical reasons; (iii) relative to the
observable matter in the universe, the cosmic phantom field can be
regarded as a cosmological source of information and negative
entropy; and (iv) in the presence of phantom radiation, quantum
level systems undergo novel radiative processes that include a
stimulated absorption phenomenon by which phantom radiation is
attenuated.

The paper can be outlined as follows. In Sec. II we consider and
extend the process of phantom energy accretion by black holes
first studied by Babichev, Dokuchaev and Eroshenko [8],
interpreting it in terms of a new thermodynamic description of a
phantom fluid which is developed and includes a negative
temperature and a positive entropy. A discussion of the properties
and meaning of the cosmological information induced by the
presence of dominating phantom energy in the universe, as well as
the implications that it may have on the emergence of life and the
anthropic principle, is included in Sec. III. By interpreting the
phantom field as a radiation field, we consider some of the
quantum statistical-thermodynamic properties of the resulting
description, including a generalized Wien spectrum (Sec. IV) and a
derivation of the generalized Planck law based on introducing
novel radiative processes (Sec. V). We finally conclude and add
some more comments in Sec. VI.

\section{Phantom-energy thermodynamics and black holes}

Phantom energy has been criticized by several authors [10-12]. The
main difficulty stemming from such criticisms concerns the issue
of phantom stability and this could still be circumvented if an
axion model is considered for the phantom field [7]. Indeed, the
large amount of papers on the phantom subject that have appeared
[4-6] before and after references [10-12] reflects the fact that
most of the criticisms can actually be regarded as manifesting
weird phantom properties that could nevertheless be accommodated
into the current and future evolution of the universe without
contradicting observations.

\subsection{Accretion of dark energy onto black holes}

Among such properties, an intriguing recent result is the
discovery by Babichev, Dokuchaev and Eroshenko [8] that all black
holes in a universe filled with a fluid violating the dominant
energy condition (i.e. a phantom-energy fluid inducing a big rip
singularity in the future)) will steadily loss all of their mass
to fully disappear, all at once, at the big rip, no matter their
initial mass or the moment at which they were formed. In fact, by
extending the classical theory of Michel [13], these authors found
that, as a result of dark energy accretion, the mass $M$ of a
black hole in a universe filled with a general quintessence scalar
field describable by means of a fluid with equation of state
$p=\omega\rho$, varies at a rate given by [8]
\begin{equation}
\dot{M}=4\pi AM^2\left(\rho+p\right) ,
\end{equation}
where $A$ is a dimensionless positive constant and $\dot{}=d/dt$.
Using then the constant equation of state $p=\omega\rho$ for the
fluid described by the quintessence field, Eq. (2.1) can finally
be re-written as
\begin{equation}
\dot{M}=4\pi AM^2(1+\omega)\rho .
\end{equation}
Thus, since in all conceivable situations $\rho>0$ should be
satisfied, accretion of dark energy onto a black hole leads the
mass of this black hole to increase if $\omega>-1$ and, as a
consequence from the negative value of the internal energy of the
phantom field, to a mass loss in the case that $\omega<-1$. The
stability of the Schwarzschild-de Sitter universe [14] to the
dark-energy accretion process is ensured by the fact that a
positive cosmological constant corresponds to $\omega=-1$ for
which the rate $\dot{M}=0$. Of course, in this case there will
still be a continuous loss of black hole mass due to Hawking
radiation, a process which is not included in the present
formalism.

For the flat geometry which our universe appears to satisfy, it
has been obtained [15] that the most general expression of the
scale factor of a universe filled with such a general quintessence
field is given by
\begin{equation}
a(t)=\left(a_0^{3(1+\omega)/2}+\frac{3}{2}(1+
\omega)t\right)^{2/[3(1+\omega)]},
\end{equation}
in which $a(0)=a_0$ is taken as the initial value of the scale
factor at the onset of the accelerating regime. This solution
satisfies the Friedmann equation for flat geometry and the
conservation law for cosmic energy. If we, moreover, re-interpret
the constant $a_0$ as the value of the scale factor at the onset
of the radiation domination era for $\omega=1/3$, then re-scaling
time $t$ so that $t\rightarrow\bar{t}=a_0^2+2t$, we in fact obtain
from Eq. (2.3) the expression of the scale factor for a
decelerating radiation-dominated universe, i.e.
$a(\bar{t})=\bar{t}^{1/2}$. Eq. (2.3) furthermore tells us that
for the interval $-1/3>\omega>-1$ the scale factor steadily
increases in an accelerating fashion, with $t$ tending to infinity
as $t\rightarrow\infty$. From the very onset of the phantom energy
regime $\omega<-1$ [3] assumed to occur at $a=a_0$, the scale
factor (2.3) predicts a super-accelerated expansion along that
regime which finally reaches the so-called big rip curvature
singularity as $t$ approaches the finite time value
\begin{equation}
t=t_*=\frac{2}{3(|\omega|-1)a_0^{3(|\omega|-1)/2}} .
\end{equation}
After the big rip, as $t>t_*$, the universe would start a period
of continuous contraction where its size tended to vanish as
$t\rightarrow\infty$. The super-accelerated expanding phase and
the contracting phase would however be disconnected to each other
because the big rip is a true curvature singularity.

If, as it is assumed throughout this paper, we set $\omega$
constant, the integration of the cosmic conservation law for
energy, $\dot{\rho}+3\rho(1+\omega)\dot{a}/a=0$ [16], in this case
leads to an expression for the dark energy density
\begin{equation}
\rho=\rho_0a^{-3(1+\omega)} ,
\end{equation}
with $\rho_0$ a constant. Thus, the energy density consistently
becomes constant for the cosmological constant case $\omega=-1$,
it steadily decreases with the scale factor for $\omega >-1$, and
rather surprisingly increases with $a$ for $\omega<-1$. Once the
scale factor $a(t)$ and the energy density $\rho(t)$ have been
obtained, the field theory associated with the dark energy fluid
through the definitions of the pressure and energy density,
$\rho=\dot{\phi}^2/2+V(\phi)$,
$p=\omega\rho=\dot{\phi}^2/2-V(\phi)$, can be solved (i.e.
expressions for $\phi(t)$ and $V(\phi)$ can be obtained). We in
this way get
\begin{equation}
\phi(t)=\phi_0+\frac{2}{3}\sqrt{\frac{\rho_0}{1+\omega}}
\ln\left(a_0^{3(1+\omega)/2}+\frac{3}{2}(1+\omega)t\right)\nonumber
\end{equation}
\begin{equation}
V(\phi)=\frac{1}{2}(1-\omega)\rho_0
e^{-3\sqrt{\frac{1+\omega}{\rho_0}}(\phi-\phi_0)} .\nonumber
\end{equation}
The phantom regime is derived when we let $\phi\rightarrow i\Phi$
and $\omega<-1$ [7]. It can be seen that both $\Phi$ and $V(\Phi)$
diverge at the big rip.

Inserting now Eqs. (2.3) and (2.5) into Eq. (2.2) and integrating
we finally obtain the black hole mass time-evolution equation [17]
\begin{equation}
M(t)=\frac{M_i}{1-\frac{(1+
\omega)M_i}{\dot{M}_0\tilde{t}}\left(\frac{t}{\tilde{t}+
\frac{3}{2}(1+\omega)t}\right)} ,
\end{equation}
where $M_i$ is the initial mass of the black hole, $\tilde{t}=
a_0^{3(1+\omega)/2}$ and $\dot{M}_0=(4\pi A\rho_0)^{-1}$. Eq.
(2.6) actually shows the same predictions as those which were
first deduced in Ref. [1]. That is: (i) for $\omega>-1$, $M$
monotonically increases with time $t$, tending to a maximum value
\begin{equation}
M_{{\rm max}}=\frac{M_i}{1-\frac{2M_i}{3\dot{M_0}\tilde{t}}} ,
\end{equation}
as $t\rightarrow\infty$. This result comes about because
quintessence fields with $\omega>-1$ have positive energy which,
when accreted onto the black hole, makes the positive mass of this
black hole to increase. (ii) If $\omega=-1$, the black hole mass
remains constant, i.e. the black hole does not accrete any energy
from vacuum simply because there then is anything like a
quintessence vacuum field, but a cosmological constant instead.
(iii) Finally, and more importantly, such as it was pointed out by
Babichev, Dokuchaev and Eroshenko [8], if $\omega<-1$, then $M$
steadily decreases with time $t$ and tends to vanish as $t$
approaches the big rip singularity at time $t=t_*$. In order for
accounting for such a highly unconventional behaviour one must
resort to the feature that the internal phantom energy is negative
definite and therefore, when it is accreted onto black holes, it
subtracts rather than adds on their total energy. Moreover, near
the big rip singularity, we have that the mass of any black hole
tends to be
\begin{equation}
M\rightarrow\frac{\dot{M}_0\tilde{t}\left(\tilde{t}-\frac{3}{2}(|\omega|
-1)t\right)}{|\omega|-1} ,
\end{equation}
which remarkably does not depend on the initial mass of the black
hole. According to Babichev, Dokuchaev and Eroshenko [8], that
result means that all black holes in a universe filled with
phantom energy will tend to be equal as the big rip is approached,
and that phantom-energy accretion prevails over Hawking radiation,
at least if quantum-gravity effects are not taken into account. In
their original derivation, Babichev et al. [8] obtained exactly
the same conclusions by using a simpler expression for the scale
factor. One could still wonder, what happens with black holes
after the big rip. According to Eq. (2.6), just after the big rip
the mass of the black holes would start increasing and tend to the
finite maximum value given by Eq. (2.7) as $t\rightarrow\infty$,
such as it happens in the case of dark energy with $\omega>-1$.
The memory of the initial mass would thus be recovered, even
though the regions before and after the big rip are mutually
disconnected.

Further generalization of these results can be achieved by
considering the case where a positive cosmological constant
$\Lambda=3\lambda$ is added to the dark energy fluid. This
situation would describe a Schwarzschild-de Sitter spacetime
embedded in dark energy. If we set a vanishing initial time,
$t_0=0$, the scale factor $a(t)$ then reads [15]
\begin{equation}
a(t)=\left(\frac{2\pi G}{3C\lambda}\right)^{1/[3(1+
\omega)]}\left(e^{3(1+\omega)\sqrt{\lambda}t/2}-
Ce^{-3(1+\omega)\sqrt{\lambda}t/2}\right)^{2/[3(1+\omega)]} ,
\end{equation}
where
\begin{equation}
C=\frac{\sqrt{\lambda+8\pi
Ga_0^{-3(1+\omega)}/3}-\sqrt{\lambda}}{\sqrt{\lambda+8\pi
Ga_0^{-3(1+\omega)}/3}+\sqrt{\lambda}} ,
\end{equation}
with $a_0$ the assumed initial value of the scale factor at the
onset of the phantom energy dominance. Since we always have
$0<C<1$, a big rip singularity for the phantom regime where
$\omega<-1$ is also predicted in the presence of a cosmological
constant. Actually, that singularity takes in this case place at a
finite time given by
\begin{equation}
t=t_*= -\frac{\ln C}{3(|\omega|-1)\sqrt{\lambda}} ,
\end{equation}
which becomes shorter as the value of the cosmological constant is
made larger. The field theory can also be solved in this case.
Following the same procedure as for the scale factor (2.3) we now
attain
\begin{equation}
\phi(t)=\phi_0+\sqrt{\frac{\rho_0}{6\pi G(1+\omega)}}
\ln\left(\frac{\sqrt{C}-e^{3(1+
\omega)\sqrt{\lambda}t/2}}{\sqrt{C}+e^{3(1+
\omega)\sqrt{\lambda}t/2}}\right)\nonumber
\end{equation}
\begin{equation}
V(\phi)=\frac{1}{2}(1-\omega)\rho_0\sqrt{\frac{3\lambda}{32\pi
GC}}\sinh^2\left(\sqrt{\frac{6\pi
G(1+\omega)}{\rho_0}}(\phi-\phi_0)\right).\nonumber
\end{equation}
Also in this case the phantom regime can be obtained by
introducing the conditions $\phi\rightarrow i\Phi$ and
$\omega<-1$. Once again both the resulting field and its potential
tend to blow up as the big rip is approached.

Using solution (2.9) we can now obtain from Eqs. (2.2) and (2.5)
that in the presence of a cosmological constant the accretion of
dark energy onto a black hole makes its initial mass $M_i$ to vary
according to the law
\begin{equation}
M(t)=\frac{M_i}{1-\frac{M_i
\sqrt{\lambda}}{\dot{M}_0}\left(\frac{e^{3(1+
\omega)\sqrt{\lambda}t}-1}{e^{3(1+\omega)\sqrt{\lambda}t}-
C}\right)} ,
\end{equation}
where now $\dot{M}_0=(1-C)/(2C\rho_0)$. Clearly, again $M$ remains
all the time equal to $M_i$ if only a cosmological constant
$\omega=-1$ is present. For $\omega>-1$, $M$ again monotonically
increases with time $t$ towards a maximum $M_{{\rm
max}}=M_i/(1-M_i\sqrt{\lambda}/\dot{M}_0)$, which also occurs at
$t=\infty$. For $\omega<-1$, Eq. (2.12) can be cast in the form
\begin{equation}
M(t)=\frac{M_i}{1+\frac{M_i
\sqrt{\lambda}}{\dot{M}_0}\left(\frac{1-e^{-3(|\omega|-
1)\sqrt{\lambda}t}}{e^{-3(|\omega|-1)\sqrt{\lambda}t}-
e^{-3(|\omega|-1)\sqrt{\lambda}t_*}}\right)} .
\end{equation}
In this case $M$ once again decreases with time $t$, tending to
vanish on the neighborhood of $t=t_*$. Thus, all the qualitative
behaviours obtained from solution (2.3) are matched when one uses
solution (2.9). The effect induced by the presence of an extra
cosmological constant is to make $M$ evolve exponentially.
Actually, it can be shown that in the phantom regime corresponding
to all existing dark-energy models, the above result is always
obtained, provided that regime will show a big rip singularity in
the finite future. In fact, tachyon like models admit [18] a scale
factor solution as given by Eq. (2.9), and therefore we obtain
again for these models result (2.13). Since generalized Chaplygin
phantom models do not show any big rip singularities [19], we are
finally left only with k-essence models with non-canonical kinetic
energy [20] for which there will be a big rip future singularity
in the phantom region and this can be defined by [21]
\begin{equation}
p+\rho=-3(1-\mu)H^2/\mu < 0 ,
\end{equation}
with $\mu$ a constant satisfying $0<\mu <1$, and $H=\dot{a}/a$ the
Hubble parameter defined from the scale factor
\begin{equation}
a\propto (t-t_b)^{-2\mu/[3(1-\mu)]} ,
\end{equation}
where $t_b$ denotes the arbitrary time at which big rip takes
place. Inserting the latter two expressions into Eq. (2.1) we
finally obtain after integration
\begin{equation}
M=\frac{M_i}{1+\frac{tM_i}{t_b \dot{M}_0(t_b-t)}} ,
\end{equation}
in which now $\dot{M}_0=3(1-\mu)/(16\pi A\mu)$. Note that Eq.
(2.16) is formally the same as the expression derived in Ref. [8]
for a differently defined $\dot{M}_0$, so confirming and
generalizing to all phantom models characterized by the existence
of a sudden big rip singularity in the finite future the result
pointed out in that reference. We finally solve the field theory
for this case. Once we have Wick rotated the field $\phi$, we get
\begin{equation}
\Phi=\Phi_0-\sqrt{\frac{4\mu}{3(1-\mu)}} \ln(t-t_b) \nonumber
\end{equation}
\begin{equation}
V(\Phi) \propto\frac{\rho_0(1+\mu)}{\mu}
e^{-\sqrt{\frac{3(1-\mu)}{\mu}}(\Phi-\Phi_0)} .\nonumber
\end{equation}
We note that in this case, whereas the phantom field diverges at
the big rip, the phantom field potential tends to vanish at that
singularity.

\subsection{Thermodynamics of a dark-energy universe}

The results on accretion of dark energy onto black holes first
obtained by Babichev et al [8] and generalized above could however
get into serious conflict with thermodynamics. First of all, one
could always assume the current existence of primordial black
holes which, at the onset of dark energy domination, could have
reached a very small mass and hence a very large temperature due
to a long process of Hawking thermal evaporation. This would, in
principle, imply that Hawking radiation effects would by now
prevail over the effects of the accretion of phantom fluid if this
accretion was characterized by a lower positive temperature [22].
On the other hand, one should also expect that if the results
derived in Subsec. II B hold, then the generalized entropy [23]
defined for the whole universe, $S+S_{{\rm bh}}$ (where $S$ and
$S_{{\rm bh}}\propto M^2$ are the entropy of the phantom fluid and
the entropy of the black hole), should be expected to decrease
with time, so violating the generalized second law [23]. In what
follows we are going to show nevertheless that none of these two
apparent difficulties actually matters because, contrary to a
recent claim [22], the temperature of the phantom fluid is
definite negative, and this will allow us to re-interpret results
from a different, "quantum" standpoint.

Lima and Alcaniz have recently proposed a general thermodynamic
theory for dark energy, according to which they argue [22] that,
whereas the temperature for the phantom energy regime is always
positive, its entropy is negative definite and that, therefore, a
cosmic scenario in which the universe is filled with phantom
energy should be ruled out. There exist however rather general
thermodynamic arguments which are valid for all available cosmic
phantom-energy models, which appear to prevent the holding of that
conclusion. Thus, if the first law of thermodynamics is assumed to
hold in every involved case, then it has been shown [24] that in a
general Friedmann-Robertson-Walker flat universe filled with a
dark energy that satisfies the equation of state $p=\omega\rho$
(with $\omega$=constant), the entropy density $s$ per comoving
volume stays always constant [24], while the temperature of the
universe turns out to be given by the expression
$T=(1+\omega)\rho_0 a^{-\omega}/(s-s_0)$, with $s_0$ an
integration constant. Thus, we can generally write for the
temperature of a universe equipped with a dominating dark energy
and equation of state $p=\omega\rho$,
\begin{equation}
{\rm T}= \kappa(1+\omega)a^{-3\omega} ,
\end{equation}
where $\kappa$ is a positive constant. The interpretation of this
result can obviously resort to the most commonly used concept of
kinetic temperature which is a measure of the average
translational kinetic energy of the system. In the case of a
general dark-energy scalar field $\phi$, the kinetic term
$\dot{\phi}^2$ can be obtained from the definition of the field
itself in terms of the pressure $p$ and energy density $\rho$, the
equation of state $p=\omega\rho$, and the expression of the energy
density derived by integrating the conservation law for cosmic
energy, i.e. Eq. (2.5). This gives
$\dot{\phi}^2\propto(1+\omega)a^{-3(1+\omega)}$. Now, an estimate
of the isotropic translational kinetic energy is $a^3\dot{\phi}^2
=K(1+\omega)a^{-3\omega}$ (with $K$ a numerical constant), which
can in fact be made the same as Eq. (2.17).

Along the present paper we shall take the general expression
(2.17) as the temperature that characterizes a universe dominated
by dark energy. We next derive from Eq. (2.17) \other
thermodynamic functions of interest. Thus, Eqs. (2.5) and (2.17)
will allow us to readily get the following generalized
Stefan-Boltzmann law for a dark energy universe,
\begin{equation}
\rho=\rho_0\left(\frac{{\rm T}}{\kappa(1+
\omega)}\right)^{\frac{1+\omega}{\omega}} .
\end{equation}
We notice that, whereas for $\omega=-1$ $\rho$ becomes a simple
constant (which actually corresponds to a cosmological constant),
for $\omega=1/3$, Eq. (2.18) consistently reduces to the usual law
for radiation, and for $0>\omega>-1$ the dark energy density
decreases with the temperature. However, for the phantom regime
where $\omega<-1$, in order to preserve $\rho$ positive, we must
necessarily take ${\rm T}<0$, which is a condition that really
directly stems from Eq. (2.17) for $\omega<-1$. It follows that
$\rho$ will always increase with $|{\rm T}|$ along the entire
phantom regime. The energy of such a regime actually is bounded
from above and allows therefore the occurrence of negative
temperatures [25]. Since on the phantom regime $\rho$ increases
with the scale factor $a(t)$ it also follows that $|{\rm T}|$
increases as the universe expands on that regime. These two
characteristics are quite surprising actually and therefore they
should be added to the increasing collection of phantom weird
properties.

Finally, a general expression for the entropy of a dark energy
universe can also be obtained by using the procedure of Ref. [22]
which in the present case leads to
\begin{equation}
S=C_0\left(\frac{{\rm T}}{1+\omega}\right)^{1/\omega}V ,
\end{equation}
where $C_0$ is a positive constant and $V$ is the volume of the
considered portion within the dark energy fluid. Thus, contrary to
the claim in Ref. [22], the entropy of a dark-energy universe is
{\it always} positive, even on the phantom regime. Actually, by
inserting Eq. (2.17) into Eq. (2.19) one in fact attains that $S$
becomes a constant when we take $V$ to be the volume of the entire
universe. For usual radiation corresponding to $\omega=1/3$,
entropy and temperature can be linked by means of the general
relationship, entropy $\propto$ energy/temperature. According to
Eq. (2.19), however, that relationship should be generalized to
read: entropy $\propto$ (1+$\omega$)energy/temperature, according
to which entropy is positive along the entire interval of
conceivable values of parameter $\omega$.

Even though it is not very common in physics and could thereby be
listed as just another more weird property of the phantom
scenario, a negative temperature is not at all unphysical or
meaningless [26]. Nuclear spin and other quantum Systems with
negative temperatures have already been observed in the laboratory
and interpreted theoretically. As referred to the case of phantom
energy, the existence of a negative temperature would imply that
the entropy of a phantom universe monotonically decreased if one
would be able to supply energy to that universe. Hence, the onset
at the coincidence time of the dominance of a phantom energy,
$\omega<-1$, universe would imply the emergence of a necessarily
"hotter" cosmic evolution regime in such a way that, if two copies
of the current universe were taken, one with positive and other
with negative temperature, and put them in thermal contact, then
heat would always flow from the negative-temperature universe into
the positive-energy universe. It could yet be argued that negative
temperature is a quantum-statistical mechanics phenomenon and
therefore cannot be invoked in the classical realm. However, the
negative temperature for the universe given by Eq. (2.17) when
$\omega<-1$ [16] can still be heuristically interpreted along a
way analogous to how e.g. black hole or de Sitter temperature can
be interpreted (and derived) without using any quantum-statistical
mechanics arguments; that is by simply Wick rotating time,
$t\rightarrow i\tau$, so that the metric becomes positive
definite, and checking that in the resulting Euclidean framework
$\tau$ is periodic with a period which precisely is the inverse of
the Hawking temperature [27]. Thus, the Euclideanized black hole
turns out to be somehow "quantized". Similarly in the present
case, the phantom regime can be obtained by simply Wick rotating
the classical real scalar field [28], $\phi\rightarrow i\Phi$,
which can be generally seen to be equivalent to rotating time so
that $t\rightarrow i\tau$, too (see Sec. III). It is in this sense
that the phantom energy universe can be also regarded as a somehow
"quantized" system and that the emergence of a negative
temperature in the phantom regime can be interpreted in a
consistent way.

Since a system with negative temperature is "hotter" than any
other systems having positive temperature (i.e. energy will always
flow from the former to the latter system), even if that positive
temperature is $+\infty$ [9,26], we see that the first of the two
thermodynamical difficulties pointed out before becomes fully
solved, no matter how high the black hole temperature could be.
Concerning the second of the above difficulties, let us notice on
the other hand that, if the universe is assumed to contain a black
hole of mass M, the entropy of the dark energy fluid will be
smaller than that is given by (2.19) for $V\propto a^3$. In this
case, the entropy of the universe can be written to be
\begin{equation}
S=C_0\kappa^{1/\omega}\left(1-\frac{V_{{\rm bh}}}{V}\right) ,
\end{equation}
where $V_{{\rm bh}}\propto M^3$ is the volume occupied by the
black hole and $V$ is the volume of the entire universe. Thus, for
$\omega>-1$, $S$ decreases and, for $\omega<-1$, $S$ increases as
dark energy is accreted onto the black hole. Moreover, after
varying $S$ with respect to $V_{{\rm bh}}$ and multiplying by $T$,
one can obtain from Eq. (2.20)
\begin{equation}
\delta S\propto \pm (1+\omega)\frac{\delta E}{T}
=-\frac{(\rho+p)\delta V_{{\rm bh}}}{T},
\end{equation}
in which $\delta E=\mp\rho\delta V_{{\rm bh}}$ and the upper sign
corresponds to $\omega>-1$ and the lower sign corresponds to
$\omega<-1$. On the other hand, if on the black-hole spacetime we
assume the time taken by a light signal to travel along the entire
hole to be $t\propto M$, we can derive from Eq. (2.1)
\begin{equation}
\delta S_{{\rm bh}}\propto (1+\omega)\frac{\delta E}{T_{{\rm
bh}}},
\end{equation}
where $S_{{\rm bh}}\propto M^2$ and $T_{{\rm bh}}\propto M^{-1}$
are the entropy and temperature of the black hole.

Let us first analyze the total balance of entropy in the case of
dark energy with $\omega>-1$. Thus, from Eqs. (2.21) and (2.22) we
see that the accretion of a given amount of dark energy onto the
black hole leads to an increase of dark energy entropy which will
exceed the corresponding decrease of black hole entropy, so
preserving the second law, provided that $T>T_{{\rm bh}}$, i.e. if
$8\pi GM_{{\rm bh}}T\geq 1$, a condition which appears to be of
general applicability along the latest accelerating evolution of
the universe. It would be in this case expected however that the
competion between dark energy accretion and black hole thermal
emission, both characterized by positive temperatures, will be
governed by usual thermodynamic laws. Thus, if initially $8\pi
GM_i M_i <1$ (with $T_{{\rm
initia}}=\kappa(1+\omega)a_0^{-3\omega}$), then thermal emission
will prevail over dark energy accretion and the black hole mass
decreased initially. That situation could nevertheless be
maintained only for a while because, during accelerating
expansion, temperature (2.17) will increase at a much greater rate
than black hole temperature did due to thermal evaporation. Thus,
after a given time, both processes would first tend to balance one
another, to inexorably allowing then for dominance of the dark
energy accretion process, leading finally to a maximum black hole
mass given by Eq. (2.7). If initially we already had $8\pi GM_i
M_i>1$, then dark energy accretion prevailed over thermal
radiation along the entire evolution, so that the mass of the
black hole steadily increased up to $M_{{\rm max}}$.

As to the case of phantom energy for $\omega<-1$, we ought to
recall that according to the Carnot equality the coupling between
a negative-temperature system and a a positive-temperature system
leads to a Carnot engine of greater than 100 percent efficiency
[29]. Thus, in the present case
\begin{equation}
F=1+\frac{T_{{\rm bh}}}{|T|} > 1 ,
\end{equation}
so that the generalized entropy $S+S_{{\rm bh}}$ ought to
inexorably decrease along phantom energy accretion, so violating
the second law. In fact, Eqs. (2.21) and (2.22) tell us that in
the phantom case the entropy $S$ increases by an amount which is
smaller than the decrease of $S_{{\rm bh}}$ provided the
reasonable condition $|T|>T_{{\rm bh}}$ is again applied. This
violation of the second law was to be expected because, as it was
mentioned earlier, negative temperatures are compatible with
observationally checked "quantum" Carnot equalities violating the
second law [29]. In any event, the implication that negative
temperatures lead through the quantum Carnot engine to violation
of the second law can be argued under several presumptions.
Although discussing these arguments is clearly outside the scope
of the present paper, we want to leave open the possibility that
they might preserve the second law even in the case being
considered.

One very remarkable property of phantom energy is worth mentioning
here. Even though most of its properties might be weird, phantom
energy seems to be able to elegantly solve one of the most
conspicuous and debated paradoxes of all the physics. In fact, if
all existing black holes will simultaneously disappear at the big
rip leaving no Hawking radiation, then the information initially
lost during formation of the black holes should somehow be
recovered. Actually, the so-called quantum coherence loss paradox,
long championed by Hawking [30] and according to which an initial
pure state is transformed into a final mixed state, during the
whole process of black hole formation and subsequent complete
evaporation, is here naturally solved in at least the
phantom-energy regime in the sense that neither Hawking radiation
nor black holes are left in the final singular state where only
the big rip singularity plus a naked black hole singularity could
take place. Thus, even in the classical case, the information
paradox appears to be solved.

\subsection{Dark energy as a radiation field}

At first sight, one could be tempted to dismiss Eq. (2.17) because
a spacetime dominated by a cosmological constant has a nonzero
temperature of the order of the Hubble constant $H=\dot{a}/a$
while, according to Eq. (2.17) the temperature of a fluid with
$\omega=-1$ vanishes. If by instance we take for the scale factor
the expression (2.3), then as one approaches the value $\omega=-1$
that expression can be reduced to $a\simeq
a_0\exp\left(a_0^{-3(1+\omega)/2}t\right)$, and hence it becomes
$a\simeq a_0\exp(t)$ at the special case where $\omega=-1$. In
such a case we have then $H=1$, while for $\omega\neq -1$ the
Hubble parameter is time-dependent and given by the expression
$H\equiv H(t)=a^{-3(1+\omega)/2}$. Thus, whereas in the first case
we have a cosmological horizon of radius $H^{-1}=1$ which is
characterized by a surface gravity $K_c=1$, no constant horizon
can be defined for any $\omega\neq -1$ in which cases every point
is subject nevertheless to an isotropic acceleration which can be
given by $q\propto -\dot{H}a^3/G\propto -(1+\omega)a^{-3\omega}$.
Each of these two cases has therefore a distinct expression for
temperature: if $\omega=-1$ the surface gravity implies a
temperature $T_{{\rm dS}}\propto H=K_c=1$ [27]; if however
$\omega\neq -1$ then every accelerating point will be bathed by a
thermal radiation at the quantum Unruh temperature [31]
$T_{\omega\neq -1}\propto q$, which is precisely the expression
given by Eq. (2.17). It follows that, even though the
Gibbons-Hawking temperature for event horizons and the Unruh
temperature for accelerating systems are conceptually equivalent
as the surface gravity is nothing but a measure of how hard you
have to accelerate to stay a given short distance away from the
horizon, de Sitter space must be endowed with a temperature
$T_{{\rm dS}}\propto H$ but not with temperature $T\propto
-\dot{H}a^3$, and those spacetimes with $\omega\neq -1$ have in
turn a temperature $-\dot{H}a^3$ but not temperature $H$. Thus,
Eqs. (2.17)-(2.19) are valid only for $\omega\neq -1$. The above
discussion makes it clear that these equations are not mere
definitions devoid of physical significance, but they all possess
a precise physical meaning.

We can finally easily convince ourselves of the relevance of
temperature (2.17) if we take for the parameter of the equation of
state the value $\omega=1/3$ which obviously corresponds to usual
radiation. In fact, if $\omega=1/3$ Eqs. (2.17)-(2.19) reduce to
$T\propto a^{-1}$, $\rho\propto T^4$ and $S\propto T^3$, which are
just the values of the temperature, energy density and entropy for
the Friedmann-Robertson-Walker universe dominated by usual
radiation. Therefore, the equation of state $\omega=p/\rho$,
together with the conservation law for cosmic energy, suffice by
themselves to determine the thermodynamic properties of a general
relativistic fluid which is assumed to dominate in a
Friedmann-Robertson-Walker universe, much as e.g. the
gravitational characteristics of black holes or de Sitter space
are now known to determine well-defined and precise laws of
thermodynamics, provided quantum theory is taken into account.

As represented by a stuff governed by the conservation law for
cosmic energy and satisfying a perfect fluid equation of state the
way we have considered so far, dark energy can be looked at as
being a generalized radiation field whose characteristics are
fixed by the value of parameter $\omega$. Thus, if $\omega>-1$
dark energy would account for a quintessence radiation, if
$\omega<-1$ it would describe a phantom radiation, and finally if
$\omega=1/3$ dark energy described a conventional radiation field
for which we obtain from Eqs. (2.3) and (2.17) that
$a\propto\bar{t}^{1/2}$ (with the time $t$ re-defined so that
$t\rightarrow\bar{t}=a_0^2+2t$ and $T\propto a^{-1}$. In field
theory in curved space usual radiation can be represented by means
of a homogeneous massless, scalar field $\Phi$ which is
conformally coupled to gravity. The action integral of the system
formed by gravity plus a scalar field $\Phi$ can be written as
\begin{eqnarray}
&&S=\frac{1}{16\pi G}\int d^4 x\sqrt{g}\left[\left(1-8\pi
G\xi\Phi^2\right)\right]\nonumber\\ &&+\frac{1}{2}\int d^4
x\sqrt{g}\left[(\nabla\Phi)^2 - 2V(\Phi)\right] -\frac{1}{8\pi
G}\int d^3 x\sqrt{h}\left(1-8\pi G\xi\Phi^2\right){\rm Tr}K ,
\end{eqnarray}
where $\xi$ accounts for the coupling between gravity and the
scalar field, and all other symbols keep their conventional
meaning. We shall use the metric
\begin{equation}
ds^2 =\frac{2G}{3\pi}a(\eta)^2\left(N^2d\eta^2 +d\Omega_3^2\right)
,
\end{equation}
where $\eta=\int dt/a$ is the conformal time, $N$ is the lapse
function and $d\Omega_3^2$ is the metric on the unit three-sphere.
For conformal coupling $\xi=1/6$ and flat geometry, the action
integral (2.24) becomes
\begin{eqnarray}
&&S= -\frac{1}{2}\int d\eta N\left[-\left(\frac{a'}{N}\right)^2
+\left(\frac{\chi'}{N}\right)^2 -\frac{8\pi
G}{3}V(\Phi)a^4\right]\nonumber\\ &&= -\frac{1}{2}\int d\eta
N\left[-\left(\frac{a'}{N}\right)^2
+\left(\frac{\chi'}{N}\right)^2 -\frac{8\pi G}{3}(1-\omega)\rho_0
a^{1-3\omega}\right],
\end{eqnarray}
in which $\chi=\sqrt{4\pi G/3}a\Phi$ is the conformal field, and
we have used the definitions of the pressure and energy density
$p=\omega\rho=L=\dot{\Phi}^2/2-V(\Phi)$,
$\rho=\dot{\Phi}^2/2+V(\Phi)$ and Eq. (2.5). The equation of
motion for $\chi$ and the constraint ($\delta/\delta N$) are given
by (in the gauge $N=1$)
\begin{equation}
\chi'' =0
\end{equation}
\begin{equation}
(a')^2-(\chi')^2-\frac{8\pi G(1-\omega)\rho_0}{3}a^{1-3\omega} =0
,
\end{equation}
with the equation of motion for $a$ yielding also Eq. (2.28). From
Eqs. (2.27) and (2.28) we obtain finally
\begin{equation}
(a')^2-M^2 -\frac{8\pi G(1-\omega)\rho_0}{3}a^{1-3\omega} =0 ,
\end{equation}
where $M$ is an arbitrary integration constant. Now, if
$\omega=1/3$, then $a=(K_1 +K_2 t)^{1/2}$, i.e. unless by a time
re-scaling, essentially the same result as that we obtained from
Eq. (2.3) for $\omega=1/3$ by assuming a minimal coupling $\xi=0$.

In terms of the conformal time $\eta$ the scale factor (2.3) for
$\xi=0$ is given by
\begin{equation}
a(\eta)=\left[\frac{3(1+3\omega)\eta}{2}\right]^{2/(1+3\omega)} ,
\end{equation}
which of course is not a solution of Eq. (2.29) for any value of
$\omega$ other than $\omega=1/3$. Thus, for $\omega=-1$ we obtain
from Eq. (2.29)
\begin{equation}
a(t)=\left(\frac{3M^2}{16\pi
G\rho_0}\right)^{1/4}\sinh^{1/2}\left(\sqrt{\frac{64\pi
G\rho_0}{3}}t\right) .
\end{equation}
Finally, at the onset of the accelerating regime $\omega=-1/3$,
\begin{equation}
a(t)=\frac{9M^2}{32\pi G\rho_0}\left(\frac{32\pi G\rho_0 t^2}{9}
-\frac{9M^2}{32\pi G\rho_0}\right)^{1/2} .
\end{equation}
If we then assume that each particular value of $\omega$
corresponds to a given radiation field, it follows that all of the
resulting fields are associated with a minimal coupling between
the scalar field and gravity, except when the state equation
parameter takes on a value $\omega=1/3$, for which case the
radiation field can be obtained by either conformally coupling or
minimally coupling the field to gravity. It is in this sense that
we shall consider in what follows that a universe dominated by
dark energy is equivalent to a universe dominated by radiation.

\section{Cosmological information and the anthropic principle}

The concepts of negative temperature and negative entropy are both
meaningless within the realm of classical thermodynamics. The
mathematical definition of entropy cannot in fact accommodate
negative values both in statistical thermodynamics and in the
Shannon theory of information, where the entropy (respectively
denoted by $S$ and $H$) is given by the conventional equivalent
formulas
\begin{equation}
S=-k_B \log P ,\;\;\; H=-k\log_2 N ,
\end{equation}
with $k_B$ the Boltzmann constant, $k$ a given constant usually
assumed to be unity, $P$ the probability of any given microstate,
and $N$ the probability of the signal given the reference class of
possible signals that could have been sent. The fact that $P$ and
$N$ cannot exceed unity makes it impossible to have finite
negative values for $S$ and $H$ in the classical theory. However,
more recent developments have shown that quantum theory can allow
the emergence of both negative temperatures and negative
entropies.

Cerf and Adami studied [32] what happens to the Shannon theory
when qubits of quantum information are accounted for. In doing
that these authors extended classical probabilities to function
describing also the quantum characteristics of the system,
including both bits and qubits, within a consistent density-matrix
description that otherwise parallels classical Shannon theory. A
rather tantalizing implication from that generalization is that
negative entropies crop up in the case of quantum systems having
no classical analog. In particular, it was shown [33] that entropy
becomes negative when two particles are quantum-mechanically
entangled, even though that does not violate special relativity.
On the other hand, the occurrence of negative absolute
temperatures has already become a little bit more conventional
[26]. These take also place in quantum systems or phenomena
without classical analog, e.g. in quantum nuclear spin systems
[34], and their existence has already been experimentally checked
[34].

We shall now contend that at least in systems which admit a
statistical thermodynamic description, negative temperature
implies negative entropy, in the following sense. Given the
specific properties of the "log" function (which actually relate
characteristics of linear evolution with characteristics of
nonlinear evolution), one can still introduce a tentative
statistical definition of the entropy function which encompasses
both positive and negative values of temperature and entropy
simultaneously. Using the formula for Carnot efficiency
\begin{equation}
E=\frac{T_2 -T_1}{T_2} ,
\end{equation}
one can actually introduce a new entropy formula
\begin{equation}
S'=k_B\log (E^{-1}) =k_B\log\left(\frac{T_2}{T_2 -T_1}\right) .
\end{equation}
We notice that Eq. (3.3) connects the statistical content of
entropy with its purely thermodynamic meaning, as it can now be
checked by considering a small variation of $T_1$ which, if $T_1
>> T_2$, by Eq. (3.3) induces a small decrease of entropy, such that
$\delta S=-k_B\delta T_1/T_1$. In any event, if we assume that
$T_2<0$, then the efficiency $E>1$ and the entropy as defined by
Eq. (3.3) becomes negative. It follows that a negative temperature
would imply a negative entropy, except in the case that the system
exchanges negative amounts of internal energy, for which case Eq.
(3.3) no longer holds and must be replaced for
\begin{equation}
S'=k_B\log (E) = - k_B\log\left(\frac{T_2}{T_2 -T_1}\right) .
\end{equation}
Eq. (3.4) would in turn predict that a negative temperature $T_2
<0$ must imply a positive entropy. That is precisely the case to
which phantom energy belongs. Or more precisely, phantom energy
must be regarded as a quantum entity with no classical analog,
which is characterized by a negative temperature given by Eq.
(2.17) and a positive entropy given by Eq. (2.19). If we would for
a moment adhere to the alternate Alcaniz-Lima view [22], then the
phantom energy would still be a quantum entity without classical
analog, but characterized now by a positive temperature and a
negative entropy. The difference with the interpretation provided
in Ref. [22] is that in the present scenario having a phantom
negative entropy is not necessarily physically meaningless due to
the essential quantum nature of the phantom fluid.

We are going next to establish in exactly what sense the phantom
fluid can be considered as a quantum entity. It is known [7] that
in order to preserve weak energy condition, $\rho>0$, the phantom
scalar field should be Wick rotated (e.g. $\phi\rightarrow
i\Phi$). As pointed out in Sec. II, we can readily check that such
a rotation is equivalent to Wick rotating the time $t$ itself
(e.g. [35] $t\rightarrow -i\tau$), while preserving the field
unchanged. Thus, in the present general dark-energy scenario, the
scalar field $\phi$ is expressed in terms of time $t$ as [7]
\begin{eqnarray}
&&\phi+\phi_0=\frac{2}{3\sqrt{1+ \omega}\ell_P}\times\nonumber\\
&&\ln\left[a_0^{3(1
+\omega)/2}+\frac{3(1+\omega)\sqrt{\ell_P\rho_0}}{2}(t-t_0)\right]
,
\end{eqnarray}
where $a_0$, $\phi_0$ and $\rho_0$ are the initial values of the
scale factor, the scalar field and the energy density,
respectively, with $\ell_P$ the Planck length. In the case that
$\omega<-1$ it can be readily seen that Eq. (3.5) can be
approximated to
\begin{equation}
\phi+\phi_0\propto i(t-t_0) ,
\end{equation}
along the entire cosmic evolution ending at the big rip. From Eq.
(3.6) we can in fact deduce that Wick rotating $\phi$ while
keeping $t$ unchanged is equivalent to Wick rotating time $t$
while keeping $\phi$ unchanged. Now, it is well known that a
Euclideanized spacetime metric would describe a somehow quantized
system [27]. It is in this sense that the cosmic phantom fluid is
a quantized entity; that is, in a way which would parallel e.g the
procedure through which the quantum temperature and entropy of
black holes can be derived by simply Wick rotating time in the
maximally extended Kruskal metric [27].

Once we have made a preliminary discussion of some quantum
properties of the cosmic phantom field, let us consider the
simplest case in which the universe is initially filled only with
phantom stuff. If we then add an elementary piece of ordinary
matter to this observationally empty universe, according to our
discussion before, the entropy would drop down, meaning that at
least a bit (or qubit) of information necessarily about the added
particle has been created and is available to potential observers
(Note that essentially no direct information can be made available
on the phantom fluid as this is not directly observable by
definition). This is exactly the opposite to what would occur if
such an elementary piece of matter with positive energy is added
to a universe filled with dark energy with $\omega>-1$, in which
case the entropy of the universe increased. The creation of
negative entropy in the case of a phantom fluid cannot be
interpreted as being due to the creation of entangled correlations
between the added particle and all "phantom quanta" as this
implied knowledge about phantom being available to potential
observers. Imagine now the onset of phantom energy domination in
our real universe. The above reasoning would lead to the idea that
a huge amount of bits and qubits of information about all
observable pieces of matter encountered by the phantom field at
the start of its domination would then be created and made
available to potential observers. This is what can be dubbed as
phantom-energy induced cosmological information. It actually calls
for the presence of observers if one appeals to a general
principle for natural economics. An important point should be
mentioned at this point. It is that once phantom energy starts
dominating in the universe, as its temperature is negative, and
hence "hotter" than anything observable, it will gravitationally
be ceacelessly accreted onto all kinds of observable matter
steadily annihilating it, all the way until the big rip at which
point all of the matter in the universe, including black holes,
will completely disappear. Neither the above available
cosmological information nor this process of all matter
annihilation would nevertheless be present if instead of phantom
energy the universe were filled with a dark energy vacuum
component having $\omega>-1$.

Another standpoint from which one can look at the question of
cosmic phantom energy is the very concept of life and its origin.
In his rather controversial book "What is Life?", which collected
a series of lectures delivered in Dublin, Schr\"{o}dinger posed [36] a
key conclusion that, a way or another, remains still inescapable
today: what a living organism feeds upon is negative entropy, and
this in its very statistical thermodynamic sense. That is
currently interpreted by considering that living beings
necessarily produce a positive supply of entropy to the universe
during their vital activities which would compensate the needed
stream of negative entropy upon themselves. However acceptable
this may be for keeping the second law alive, it goes without
solving the key question: for an essential element of life to make
its first appearance in the universe, the site where it appeared
ought to be prepared to provide the necessary and immediate supply
of negative entropy for the incipient life element to continue, as
in this case "egg would precede hen". Thus, since any living
organisms are made up of positive energy, while original life
elements could not be consolidated if the universe was then filled
with any form of dark energy with $\omega\geq -1$, it can
perfectly do so in a universe filled with phantom energy, as in
this case the very act of the organism appearance does imply a
supply of negative entropy. On the other hand, as discussed
before, an universe filled with phantom energy is also prepared to
be observed. It is in these senses that it could be thought that
any form of anthropic principle [37] had to be formulated in terms
of the emergence of an epoch when the universe started to be
dominated by phantom energy.

\section{The Wien spectrum}

We shall apply in this and the next sections some aspects of the
thermodynamic theory for $\gamma$-fluid developed by Lima and Maia
[38] to the case of a general quintessential model of dark energy
and, in particular, to the case of phantom energy. According to
the discussion in Sec. II we can assume that the dark-energy field
corresponds to a kind of radiation field with a Wien-type spectrum
given by
\begin{equation}
\rho_{T}(\nu)=\alpha\nu^{\beta}\phi\left(\nu T^{\lambda}\right),
\end{equation}
where $\alpha$ is a positive constant, $\nu$ is the frequency, $T$
is absolute temperature, and the parameters $\beta$ and $\lambda$
will be determined by imposing the two following constraints to be
satisfied by Eq. (4.1)
\begin{equation}
\rho(T)=\int_0^{\infty}\rho_T(\nu)d\nu\propto
T^{(1+\omega)/\omega}
\end{equation}
\begin{equation}
N(T)=\int_0^{\infty}\frac{\rho_T (\nu)d\nu}{\nu}\propto
T^{1/\omega} ,
\end{equation}
which respectively define the energy density and the particle
number density. Following Lima and Maia [38], let us then define a
new variable $u=\nu T^{\lambda}$, so that the constraints (4.2)
and (4.3) can be re-cast as
\begin{equation}
\rho(T)=\frac{\alpha}{T^{\lambda(1+\beta)}}\int_0^{\infty}
u^{\beta}\phi(u)du\propto T^{(1+\omega)/\omega}
\end{equation}
\begin{equation}
N(T)=\frac{\alpha}{T^{\lambda\beta}}\int_0^{\infty}
u^{\beta-1}\phi(u)du\propto T^{1/\omega} ,
\end{equation}
where we have used Eq. (4.1). From Eqs. (4.4) and (4.5) we can
finally deduce that $\lambda=-1$ and $\beta=1/\omega$ and,
therefore, our generalized Wien law is given by
\begin{equation}
\rho_T(\nu)=\alpha\nu^{1/\omega}\phi\left(\frac{\nu}{T}\right) ,
\end{equation}
which, as expected, reduces to the known Wien law for blackbody
radiation if we set $\omega=1/3$.

We note then that the temperature should not appear as simply $T$
but always in the combination $T/(1+\omega)$ (see Sec. II), both
in the energy density $\rho_T(0)$ and in the energy density
$\rho(T)$. We can therefore write the Wien spectrum (4.1) for the
phantom energy case as
\begin{equation}
\rho_{T}(\nu)=\alpha\nu^{\beta}\phi\left(\nu |T|^{\lambda}\right),
\end{equation}
and hence using the same procedure as before, we recover the same
Wien law, but referred to absolute value of temperature, also for
a phantom radiation, i.e. for $\omega<-1$ and $T<0$,
\begin{equation}
\rho_T(\nu)=\alpha\nu^{1/\omega}\phi\left(\frac{\nu}{|T|}\right) .
\end{equation}
This actually gives the most general expression for the Wien law
and is valid for any positive and negative value of the parameter
$\omega$.

\section{Spontaneous and stimulated absorption}

Let us consider a set of two level systems characterized by the
resonant frequency $\nu$, immersed in phantom radiation at
negative temperature $T<0$. Assuming the simple dipole-moment
approximation we label the energy levels of the system by $n$ and
$m$ (see Fig. 1), with $E_m -E_n=h\nu$. Following a line of
reasoning analogous to that which led Einstein to introduce his
celebrated absorption and emission coefficients [39], we can now
assume the occurrence of novel radiative processes if the
radiation field is made up of a substance characterized by an
equation of state with $\omega<-1$. In that framework the
probability that a system is in the energy level $E_i$ ($i=n,m$)
is given by
\begin{equation}
W_i = p_i e^{E_i/(k_B|T|)} ,
\end{equation}

\begin{figure}
\includegraphics[width=.9\columnwidth]{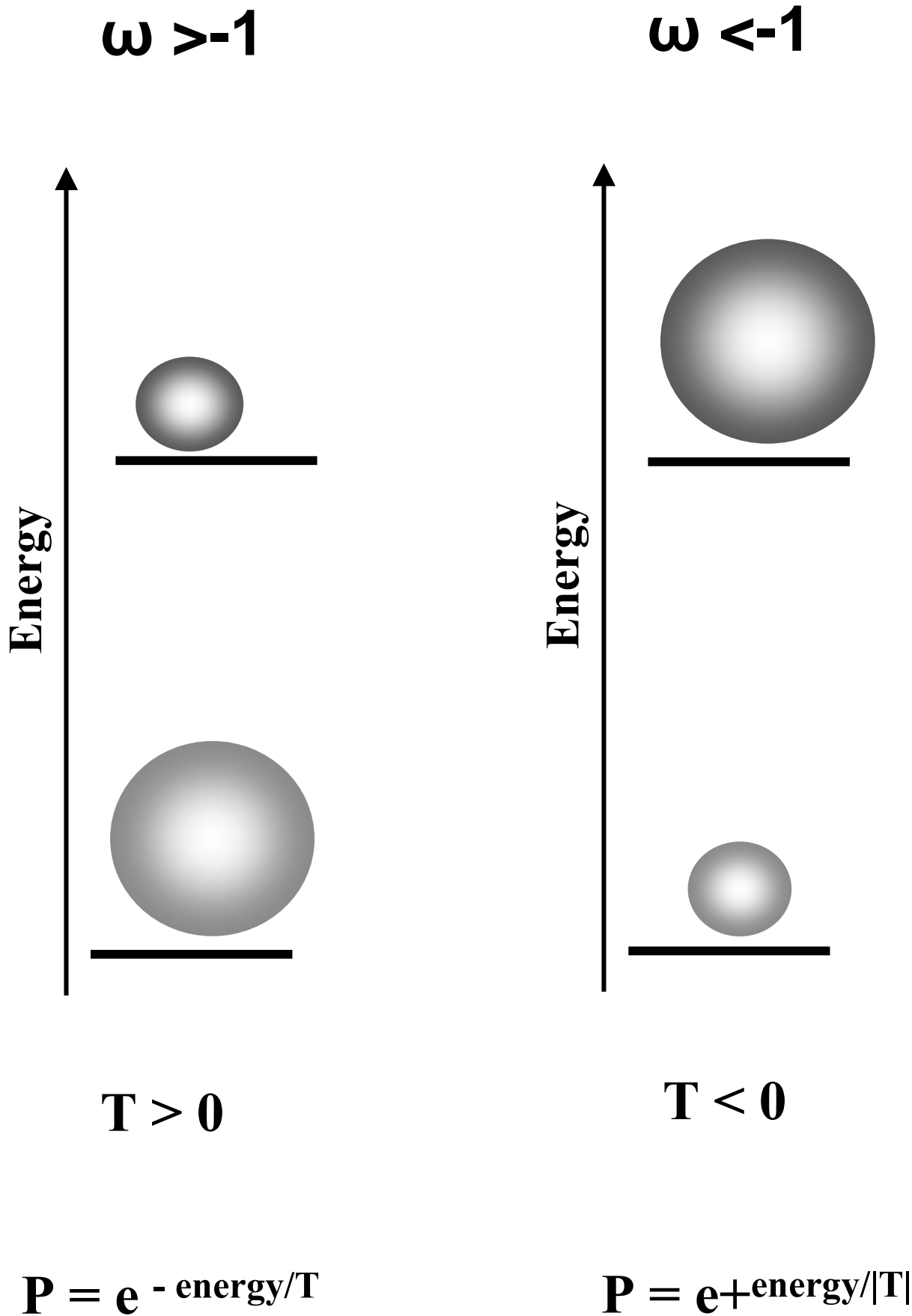}
\caption{\label{fig:epsart} Population of given energy levels in
the cases that: (left) $\omega>-1$ and hence $T>0$, and (right)
$\omega<-1$ and hence $T<0$. The area of the circles over the
energy levels would qualitatively be proportional to the
population.}
\end{figure}

\noindent where $K_B$ is the Boltzmann constant and $p_i$ is the
temperature-independent statistical weight of the {\it i}th state.
Now, due to the presence of phantom energy which would only
exchange discrete amounts of negative internal energy
$(1+\omega)\rho\Delta V$ with the level systems, there will be
three new Einstein-like coefficients, here denoted by $B_m^n$,
$B_n^m$ and $A_n^m$ (See Fig. 2). $A_n^m$ will correspond to a
spontaneous absorption coefficient which would be evaluable from
first principles using quantum mechanics and happens in the
absence of any phantom radiation, but creates quanta of phantom
energy; $B_m^n$ is a new induced emission coefficient and $B_n^m$
is a new stimulated absorption coefficient. The latter two
coefficients would only occur in the presence of phantom
radiation. These three coefficients will be related with the
following rates of probability transition.
\begin{equation}
\dot{W}_{mn}=B_m^n\rho_T(\nu)
\end{equation}
\begin{equation}
\dot{W}_{nm}=B_n^m\rho_T(\nu)
\end{equation}
\begin{equation}
\dot{W}_{nm}=A_n^m
\end{equation}

Thus, an equilibrium condition can be introduced in the considered
system which reads
\begin{equation}
p_m
e^{E_m/(k_B|T|)}B_m^n\rho_T=p_ne^{E_n/(k_B|T|)}\left(B_n^m\rho_T
+A_n^m\right) ,
\end{equation}
and if the approximation $|T|>>0$ is applied, then $p_m
B_m^n\simeq p_n B_n^m$, and hence
\begin{equation}
\rho_T(\nu)=\frac{A_n^m}{B_n^m\left(e^{h\nu/(k_B|T|)}-1\right)},
\end{equation}
where we have used $E_m-E_n=h\nu$. Comparing with the generalized
Wien law (4.8), we finally obtain a generalized Planck law, with
[34]
\begin{equation}
A_n^m=\alpha B_n^m \nu^{1/\omega} ,
\end{equation}
\begin{equation}
\phi(\nu/|T|)=\frac{1}{e^{h\nu/(k_B|T|)}-1}
\end{equation}
being the average occupation number, and finally
\begin{equation}
\rho_T(\nu)=\frac{\alpha\nu^{1/\omega}}{e^{h\nu/(k_B|T|)}\pm 1} ,
\end{equation}
where the case of fermions has been also included. The law (5.9)
is the most general Planck law and can be applied to radiation
characterized by an equation of state $p=\omega\rho$, where
$\omega$ can take on any positive or negative value.

\begin{figure}
\includegraphics[width=.9\columnwidth]{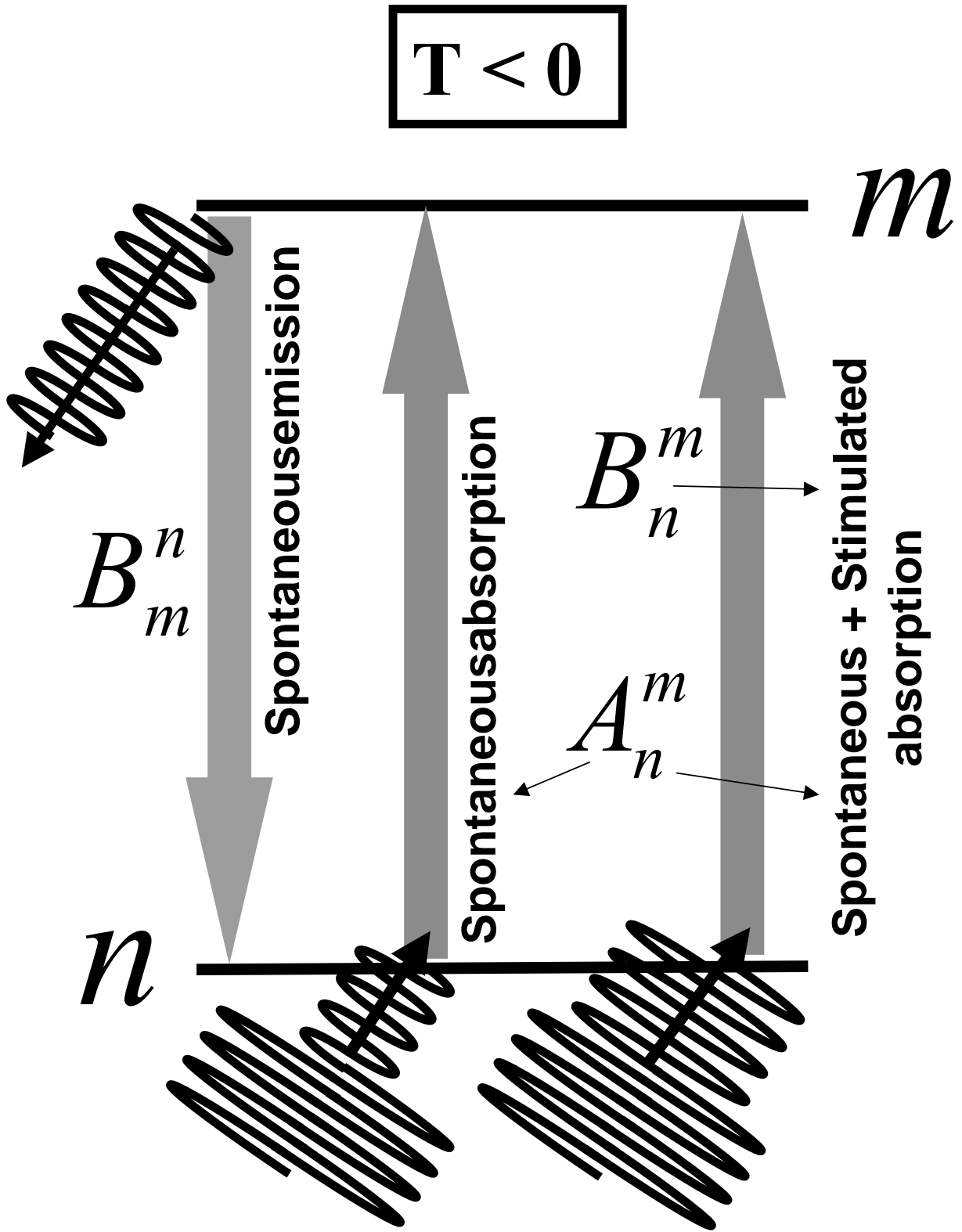}
\caption{\label{fig:epsart} Pictorial representation of the
possible radiative processes induced by phantom radiation on a
two-level system as expressed in terms of the Einstein-like
coefficients: $A_n^m$ for spontaneous absorption, $B_m^n$ for
induced absorption, and $B_n^m$ for stimulated absorption. No
stimulated emission can take place in the presence of phantom
energy with negative temperature.}
\end{figure}

As a test of consistency we finally calculate e.g. $\rho(T)$ from
Eq. (5.9) for the fermion case. We in fact obtain
\begin{eqnarray}
&&\rho(T)=\alpha\int_0^{\infty}\frac{\nu^{1/\omega}d\nu}{e^{h\nu/(k_B|T|)}+1}=
\alpha\left(\frac{k_B|T|}{h}\right)^{(1+\omega)/\omega}\times\nonumber\\
&&\left(1-
2^{-1/\omega}\right)\Gamma\left(\frac{1+\omega}{\omega}\right)\zeta\left(\frac{1+
\omega}{\omega}\right) ,
\end{eqnarray}
where $\Gamma$ and $\zeta$ are the Gamma and zeta functions [40],
respectively. That is the result what was to be in fact expected.

Before closing up this section we would like to briefly comment on
the effect that negative phantom temperatures may have on laser
effect. In fact, it is well known that laser theory is based on
the occurrence of stimulated emission which, together with
population inversion, leads to the necessary coherent radiation
amplification leading to the laser operation. In the case of
negative phantom temperature, even though population inversion is
guaranteed, there is anything like a stimulated emission and
lasing can never be produced. What one could instead suppose is
the possibility for a contrary or anti-lasing process. If in our
negative temperature system one would restore a larger population
in the lower energy level by the use of an appropriate procedure,
then stimulated absorption would produce a damping, rather than
amplification, of the coherent radiation. The construction of such
an anti-laser, or more properly "lasar" (light attenuation by
stimulated absorption of radiation), device could be of some
usefulness in atomic and molecular physics. On the other hand,
stimulated absorption processes might provide a clue for
explaining the lack of phantom-energy radiative processes in
cosmology.

\section{Conclusion and further comments}

There have been two recent important contributions to the question
of the thermodynamics of dark energy. On one hand, Alcaniz and
Lima [22] derived general dark energy thermodynamic expressions
that, when extended to the phantom regime, implied positive values
for its temperature and negative values for its entropy. By using
two independent compelling arguments, here it is nevertheless
shown that what becomes negative in the phantom regime is
temperature, while entropy is kept definite positive along the
entire range of negative state equation parameter. In any event,
since classically having either a negative entropy or a negative
temperature, or both, makes no physical meaning, the first obvious
conclusion seems to be that phantom energy cannot exist. This
situation is somehow reminiscent to the one which was posed in the
mid-seventies when it was first realized that the black hole
entropy had to be finite [41]. Hawking himself has many times
stressed [42] that, at that time, it seemed that since black hole
cannot emit anything classically, such a conclusion was
meaningless, too. The solution to this paradoxical situation was,
as it is now widely acknowledged, to appeal to the quantum nature
of black holes which allowed them to radiate thermal particles. We
argue in this paper that the solution to the above apparent
phantom thermodynamic paradox is again by appealing to the
essential quantum nature of the phantom stuff. In fact, if, in
spite of by definition violating the dominant energy condition,
i.e. $p+\rho<0$, phantom energy is consistently assumed to satisfy
the weak energy condition, $\rho>0$, then it can be interpreted
that the phantom spacetime becomes Euclideanized and this is known
to provide the natural framework where quantum temperature and
quantum entropy can be consistently defined [27]; i.e. the cosmic
phantom fluid can be essentially regarded as a quantum system
having not any classical analogs. Now, negative temperatures (or
even entropies) can perfectly exist in the quantum realm. It is in
this way that the above thermodynamic phantom paradox is solved in
the present paper. Our conclusion thus is that there can exist a
cosmic quantum field characterized by a temperature that is always
negative and a positive definite entropy, and that, moreover, if
the universe would happen to be currently dominated by such
phantom energy, then this would introduce a universal element of
cosmological information that would make every piece of observable
matter to be really observable and allow the very existence of
living organisms in the universe. On the other hand, when
considering accretion of phantom energy onto black holes,
Babichev, Dokuchaev and Eroshenko have showed [8] that the mass of
the black holes undergo a gradual decrease and tend all to zero at
the big rip. In this work we confirm and extend that result,
consistently interpreting it in terms of the above conclusion on
phantom thermodynamics. This realization is important in at least
two respects. It firstly implies that once black holes are formed
before or after phantom energy domination, all known evolution
processes of existing black holes become inexorably dominated by
phantom energy accretion and finally subject to what could be
dubbed as a "democracy before death" principle by which such black
holes are all equalized before disappearing all at once at the big
rip, no matter their initial mass or the time when they were
formed. The reason of such a domination of phantom energy
accretion (which actually extends over any emission processes of
any observable matter system) essentially resides on the fact
that, however small its absolute value may be, a negative
temperature is always hotter than any positive temperature, even
if this is infinite. Secondly, it is also a conclusion of the
present paper that, since the phantom-induced annihilation process
prevails always over Hawking thermal emission, during the whole
black hole evolution process leading from formation to final
disappearance, quantum coherence is preserved, as in this case
neither the black hole nor any thermal radiation emitted from it
is left in the final state. This offers a rather comfortable
solution to the so-called black-hole information paradox [27] and
becomes still another positive consequence from the existence of
phantom energy in the universe.

The formulation of some main theoretical basis for establishing
the quantum statistic thermodynamics of phantom energy has been
also aimed at in this work. By assuming that any form of dark
energy can be taken to be a radiation field, we have thus
considered a generalized Wien law and hence a generalized Planck
radiation law. These laws turn out to be described by the same
general expressions as for the usual dark energy case first
derived by Lima and Maia [38], but referred to the absolute value
of phantom temperature. This conclusion comes about as a
consequence from the necessary introduction of novel Einstein
coefficients and the different probability law for the case of
phantom energy. The extra Einstein coefficients correspond to new
radiative processes that include a stimulated absorption
phenomenon, which would attenuate the intensity of the phantom
radiation.

It will only be by collecting more cosmological data that we will
be able to finally decide on whether or not phantom energy is the
form of dark energy that currently operates in the universe to
drive its observed accelerating expansion. However, even in the
event that other kind of dark energy turned out to be the favoured
stuff dominating the current universe, there could still be
sufficient room for phantom energy to dominate over all other
cosmic stuffs during other epochs along the universal evolution,
including the primordial inflationary period, such as has been
recently suggested [43]. Anyway, we hope that the contents of the
present paper may help to convince cosmologists that, whatever the
final conclusion on the above subjects may be, phantom energy is a
physical concept with sufficient theoretical interest by itself as
to pursue active research on it.

\acknowledgements

\noindent We thank Professors Lima and Babichev for useful
explanations, discussions and correspondence. This work was
supported by DGICYT under Research Project BMF2002-03758. The
authors wish to respectfully dedicate the present work to the
memory of all those that were killed, wounded or directly affected
in the terrorist attacks of March 11 in Madrid


\begin{references}
\bibitem {28} One can in principle Wick rotate the field either
clockwise, $\phi\rightarrow i\Phi$, or anti-clockwise,
$\phi\rightarrow -i\Phi$ (see Ref. [31]). It will be seen in Sec.
III that $\phi\rightarrow i\Phi$ will be equivalent to rotating
time anti-clockwise and $\phi\rightarrow -i\Phi$ will be
equivalent to rotating time clockwise.
\end{references}
\end{document}